\documentclass[epj,final]{svjour}
\usepackage{graphics}
\usepackage{epsfig}
\newcommand{\bm}[1]{ \mbox{\boldmath $#1$}  }

\begin{document}
\title{Structure of low-lying $^{12}$C-resonances}
\author{R. \'Alvarez-Rodr\'{\i}guez\inst{1} \and E. Garrido\inst{2}
\and A.S. Jensen\inst{1} \and D.V. Fedorov\inst{1} \and H.O.U. Fynbo\inst{1}
}                     
%
%
\institute{Institut for Fysik og Astronomi, Aarhus Universitet, 
DK-8000 Aarhus C, Denmark \and Instituto de Estructura de la Materia.
Consejo Superior de Investigaciones Cient\'{\i }ficas. \\
Serrano 123, E-28006 Madrid, Spain}
\date{Received: date / Revised version: date}
%
\abstract{
The hyperspherical adiabatic expansion is combined with complex
scaling and used to calculate low-lying nuclear resonances of $^{12}$C
in the  $3\alpha$-model.  We use Ali-Bodmer potentials and compare
results for other potentials $\alpha-\alpha$ with similar
$^{8}$Be-properties.  A three-body potential is used to adjust the
$^{12}$C-resonance positions to desired values extending the
applicability of the method to many-body systems decaying into three
$\alpha$-particles.  For natural choices of three-body potentials we
find $14$ resonances below the proton separation threshold, i.e. two
$0^{+}$, three $2^{+}$, two $4^{+}$, one of each of $1^{\pm}$,
$2^{-}$, $3^{\pm}$, $4^{-}$, and $6^{+}$.  The partial wave
decomposition of each resonance is calculated as function of
hyperradius. Strong variation is found from small to large distance.
Connection to previous experimental and theoretical results are
discussed and agreements as well as disagreements are emphasized.
} 
\maketitle

\section{Introduction}

The low-lying nuclear bound states and resonances of $^{12}$C have
been the subject of numerous investigations.  Surprisingly a number of
issues are still not settled, e.g. what are the energies of the
low-lying resonances below the proton separation threshold 15.96~MeV
of excitation energy, what are their angular momenta, their structure
in general, and their decay properties.  Substantial experimental
efforts are presently devoted to find answers to these questions
\cite{dig06}.  This is partly motivated by the interest from
astrophysics \cite{fyn05}, but the research has basic intrinsic
interest in its own right.

Experiments often focus on specific properties and/or specific
resonances, but taken together more and more accurate data accumulate
\cite{ajz90}. Open questions remain on 0$^+$ and 2$^+$ resonances 
\cite{kur04} above the first 0$^+$ resonance at 7.6 MeV, e.g. in 
\cite{ajz90} there are only tentatively assigned candidates. Recently 
this has been addressed experimentally both by inelastic scattering of
alpha-particles on $^{12}$C \cite{joh03,ito04}, and by studies of the
beta-decays of $^{12}$N and $^{12}$B \cite{dig05}. These studies are
in reasonable agreement on the existence of a broad 0$^+$ state near
10~MeV, but give three different suggestions for the position and
width of the first 2$^+$ resonance.

The accurate measurements and the related careful analysis call for a
reliable theoretical description.  The theoretical efforts go back at
least fifty years but consensus has not been reached.  Advanced recent
attempts can be divided into microscopic models dealing either
directly with an unbiased $12$-body nuclear problem or with various
cluster constraints of four nucleons within each of the three
$\alpha$-particles.  Within the first group we find the no-core shell
model \cite{nav00}, the stochastic variational method \cite{mat04} and
Greens function Monte Carlo calculations \cite{pie05}. None of these
methods have so far been able to provide satisfactory answers to the
present continuum problem.  In the second group we find antisymmetric
molecular dynamics \cite{kan98}, Fermionic molecular dynamics
\cite{nef04}, and resonating group methods \cite{des02}.  These methods 
are very ambitious as well but more directed to account for possible
cluster structures. However, they have also not provided convincing
answers as for example evidenced by mutual disagreements.
Furthermore, the same tendency of concentrating on specific properties
and resonances is seen as in the experimental studies.  Systematic
investigations of all low-lying resonances with a given model are very
few \cite{ueg77,ueg79,bij99,kan98}, and not easily reconciled
with the newest experimental information.

These microscopic theories have so far not provided energy
distributions of the $\alpha$-particles after decay of the resonances.
These observables are difficult to compute with a reasonable accuracy,
but nevertheless this is precisely where the accurate and complete
experimental information accumulates \cite{dig05}.  To address these
problems we scale down the ambitions to a practical yet challenging
size where systematic investigations are possible.  $-3\alpha$
cluster models suggest themselves since energies below the proton
separation threshold at 15.96~MeV only allow $3\alpha$ and
$\gamma$-emission.  Many-body resonances decaying into three particles
necessarily reduce to a three-body problem at large distance. The
validity of $3\alpha$ models extends for short-range interactions to
surprisingly small distances where the three particles almost touch
each other \cite{fed96,fed02,fed03}.  The $\alpha$-cluster structure
may dominate and the description can possibly be extended to even
smaller distances.  However, detailed information at small distance
can not be expected without introduction of nucleonic degrees of
freedom.  For this reason three$-\alpha$ models can only be expected
to describe intermediate and large-distance properties related to
$\alpha$-clustering.  In particular electromagnetic transitions depend
sensitively on short-distance properties of the wavefunctions and are
therefore in general not reliably accessible within these models.

The limitations of these models are not established. The energy
distributions after decay are large-distance properties and as such
possible to compute rather accurately with few-body techniques
\cite{gar06}. However, even this $3\alpha$ problem is very challenging. 
First because we are then dealing with a three-body quantum mechanical
Coulomb problem in the continuum.  Second because we want a
comprehensive picture describing all resonances and all three-body
observables within the same model.  The second requirement is
especially demanding since the resonance may vary from two-body to
genuine three-body large-distance structures.  Coulomb and short-range
interactions must then be treated on an equal footing.  However, the
techniques and methods are available and recently applied in similar
practical three-body computations \cite{gar06b}. 

We shall use the hyperspherical adiabatic expansion method
\cite{nie01} combined with complex scaling of the coordinates \cite{fed03}.
This method restricts the applications to resonances where the width
is relatively small compared to the real part of the energy. In any
case as these states are closest to the real axis they have the
strongest direct influence on observables.  Furthermore, as obtained
by analytical continuation they have the smallest uncertainty due to
the choice of parametrization of the interactions.  To describe
properly the different types of asymptotic large-distance structures
we use the Faddeev decomposition.  Then the individual components can
simultaneously account for several two-body substructures like narrow
resonances in corresponding subsystems.  This is a tremendous
advantage over one-component methods.

The purpose of the present work is to (i) establish that the
$3\alpha$-cluster model can answer questions related to properties of
many-body resonances, (ii) give a survey of the structures of possible
$^{12}$C-resonances below the excitation energy of 15.96~MeV, and (iii)
lay out the foundation for calculations of energy distributions after
decay.  We shall build on experience gained from three-body
investigations of bound states and continuum properties of $^{6}$He
\cite{gar05a}, $^{11}$Li \cite{gar02a,gar02b,gar06}, $^{17}$Ne
\cite{gar03}, $^{6}$Be and $^{6}$Li \cite{gar06a} as well as more
general investigation of three-body resonance properties
\cite{gar04a,gar04b,gar05b}.  In section 2 we shall first give a brief
sketch of the theoretical formulation.  In section 3 we discuss the
resonances and their structures for a specific two-body interaction.
In section 4 we describe the sensitivity to interactions.  Finally
section 5 contains a summary and the conclusions.

\section{Basic theoretical ingredients}

We use techniques described in details in previous publications. It
suffices here to give a brief sketch of the procedure employed in the
present work. The general framework can be found in \cite{nie01}.  The
new results found for $^{12}$C will be discussed in more details in
the following sections.  The adiabatic hyperspherical expansion method
with the Faddeev decomposition is well established in applications to
nuclear three-body systems \cite{gar02a}.  The combination with
complex rotation to compute resonances is also known to be very
efficient for these systems \cite{fed03}. The method consists of a
number of steps: First we define the coordinates where the hyperradius
$\rho$ is the most important, i.e.
\begin{eqnarray}  
    \rho^{2} =  \frac{4}{3} \sum^{3}_{i<j}
 \left(\bm{r}_{i}-\bm{r}_j\right)^{2} = 
  4  \sum^{3}_{i=1} 
 \left(\bm{r}_{i}-\bm{R}\right)^{2}  \label{e120} \;,
\end{eqnarray}
where $\bm{r}_{i}$ is the coordinate of the $\alpha$-particle number
$i$ and $\bm{R}$ is the center-of-mass coordinate of $^{12}$C.  The
factor of $4$ is arbitrary and chosen to correspond to a normalization
mass equal to that of the nucleon.  The remaining relative coordinates
are all dimensionless angles.

Second we choose a two-body interaction as input for the angular
part of the complex rotated Faddeev equations.  This interaction
should reproduce the low-energy scattering properties of all pairs of
particles in the three-body system. A smaller amount of data like
scattering length and effective range may also be sufficient, or
perhaps the low-lying two-body resonance energies and their widths.
We use the Ali-Bodmer potentials \cite{ali66} in various combinations,
and sometimes compared with results from even simpler interactions
adjusted to reproduce selected $^{8}$Be-properties.  The hyperradius
is fixed and the solutions to the angular wavefunction $\Phi_n$ are
calculated.

Third a three-body Gaussian potential, $V_{3b}= S
\exp(-\rho^2/b^2)$, is selected.  Such potentials are necessary to 
reproduce accurately few-nucleon observables \cite{are04}.  We choose
$b \approx 6$~fm such that $\rho=b$ corresponds to three touching
$\alpha$-particles. This three-body potential is diagonal, i.e. added
to each of the adiabatic potentials whereas we assume that the
corresponding three-body couplings are zero.  This construction
maintains the structure of a three-body state, but by varying the
strength the energy position can be adjusted to reproduce the measured
value for each resonance.  

An individual adjustment is crucial in a comparison of partial widths
which depend exponentially on the energy due to barrier penetration.
Since the small-distance structure easily can vary from state to state
the three-body potential should in principle be state-dependent.
However, we choose the same potential for all states of given angular
momentum and parity, $J^\pi$.  One simple reason is that this
essentially corresponds to the number of known 
experimental data and we can then
estimate or predict the positions and widths of other states with the
same $J^\pi$. States with different $J^\pi$ can be expected to differ
more in structure and thus require different three-body potentials.
We choose to ignore the two bound states because
they, and especially the ground state, have a smaller size and
their structure can then be expected to involve more nucleonic (less
$\alpha$-cluster) degrees of freedom.  Although it is possible to
construct a three-body potential which reproduces energies of both the
ground state and the lowest $0^+$ resonance \cite{fet05,fet04} it is
uninteresting in the present context since the
important intermediate and large distance-structure would remain
unchanged.

Fourth  we find the radial wavefunctions, $f_n (\rho)$, by solving
the coupled set of radial equations arising from expansion of the
total wavefunction $\Psi$ on $\Phi_n$, i.e.
\begin{equation} \label{e125}
\Psi^{(JM)} = \frac{1}{\rho^{5/2}}\sum_n 
 f_n (\rho) \Phi_{n}^{(JM)} (\rho,\Omega)\;.
\end{equation}
The expansion coefficients $f_n (\rho)$ are exponentially decaying for
resonances when the rotation angle $\theta$ of the hyperradius is
larger than corresponding to the three-body resonance.  Then both real
and imaginary parts, $E_0=E_R-iE_I$, of the resonance energy, $E_0$,
are determined by $f_n(\rho \to \infty) = C_n \exp(+i\kappa\rho)$ with
$\kappa = \sqrt {2mE_0/\hbar^2}$.

In principle some of the channels could correspond to two-body
bound states or narrow resonances which asymptotically would have
radial wavefunctions decaying with a wave number corresponding to the
three-body energy minus that tied up in the two-body system.  In
practice this only has marginal importance since the radial
wavefunction in any case decreases towards zero for large
distances. The interpretation, and maybe the analysis, in terms of
direct or sequential decay may however be very different.

Fifth the structures of the resonance wavefunctions are computed and
expressed in terms of two-body partial waves.  We extract the
amplitudes (above called the radial wavefunctions) for each of the
adiabatic potentials, and then we partial wave decompose each of these
adiabatic components.  The dependence on hyperradius can be
substantial corresponding to a dynamical evolution of the resonance
from small to large distances.

In these steps the problems with the Coulomb interaction are not
mentioned although the large-distance asymptotic behavior is
mathematically unknown for continuum states. We adopt the pragmatic
procedure to treat the Coulomb interaction completely numerically.
The general behavior and convergence properties can then be
investigated and extracted if necessary \cite{gar06b}.  Problems would
reveal themselves intrinsically in the process as numerical
inconsistencies.

\section{Resonances and their structure }

The sequence of resonances is computed with the Ali-Bodmer potential
``AB(a')'' \cite{ali66} which reproduce the $s$, $d$ and qualitatively
also the $g$-wave $\alpha$-$\alpha$ phase shifts.  The interactions
for higher partial waves are the same as for $g$-waves.  The $^{8}$Be
resonances of $0^+$, $2^+$, $4^+$ have energies ($E_R$) and widths
($\Gamma = - 2 E_I$) given by $(E_R,\Gamma)=(100,0.010)$~keV,
$(2.7,1.5)$~MeV, and $(9.7,8.0)$~MeV, respectively.  The definitions
correspond to poles of the $S$-matrix, and should be compared to the
experimental values of $(E_R,\Gamma)= (92,0.0068)$~keV,
$(3.1,1.5)$~MeV, and $(11.5,4)$~MeV, respectively.  The $4^+$-state is
not very well reproduced although the information is derived from
phase shifts reproducing the data.

\subsection{Partial waves}

For each of the $14$ resonances angular momentum and parity constrain
the contributing two-body components within the three-body system.
The partial wave orbital angular momenta between two particles in one
Jacobi system and that of the third particle are denoted by $\ell_x$
and $\ell_y$, respectively.  The number of partial waves is rather
small because the $\alpha$-particle has zero spin and the symmetry
requirement eliminates in addition several components.  The Faddeev
components are the same in all three Jacobi systems.  The number of
basis states needed to describe the necessary number of partial waves
can only be decided a posteriori.  Accuracy is optimized by choosing
large basis sets, enumerated by the hyperspherical quantum number
$K_{max}$, when the contribution is large for a given partial wave.

\begin{table}
\vspace*{-0.1cm}
\caption{\label{tab1}
Components included for each $J^{\pi}$ state of $^{12}$C. The columns
two and three give orbital angular momenta in each Jacobi system
column four gives the maximum value of the hypermomentum $K$. In the
last columns $W_i$ give the probabilities in \% for finding these component
in the $i$-th resonance.  }
\begin{footnotesize}
\begin{center}
\begin{tabular}{|cccc|ccc|}
\hline
$J^\pi$& $\ell_x$ & $\ell_y$  & $K_{max}$ & $W_1$ & $W_2$ & $W_3$  \\
\hline
$0^+$ & 0 & 0 & 180 & 83  & 44  &\\
        & 2 & 2 & 180 & 16 & 54 &\\
        & 4 & 4 & 80  & 0 & 2 & \\
\hline
$1^+$ & 2 & 2 & 180 & 86 & &\\
      & 4 & 4 & 180 & 14 & &\\
\hline
$2^+$ & 0 & 2 & 120 & 45  & 13  & 4  \\
      & 2 & 0 & 120 & 45  & 10  & 45 \\
      & 2 & 2 & 180 & 7  & 65  & 5 \\
      & 2 & 4 & 80  & 1 & 0 & 32   \\
      & 4 & 2 & 100 & 0 & 2 & 14 \\
\hline
$3^+$ & 2 & 2 & 90 & 36 & & \\
      & 2 & 4 & 90 & 32 & & \\
      & 4 & 2 & 90 & 32 & & \\
\hline
$4^+$ & 0 & 4 & 80  & 27 & 5 & \\
      & 2 & 2 & 120 & 39 & 8  &  \\
      & 2 & 4 & 160 & 2  & 44 & \\
      & 4 & 0 & 120 & 27 & 18  &\\
      & 4 & 2 & 120 & 4 & 25 & \\
      & 4 & 4 & 100 & 0 & 1  & \\
\hline
$6^+$ & 0 & 6 & 50 & 3 & & \\
      & 2 & 4 & 90 & 62 & & \\
      & 2 & 6 & 50 & 1 & & \\
      & 4 & 2 & 70 & 15 & & \\
      & 6 & 0 & 70 & 19 & & \\
\hline
$1^-$ & 0 & 1 & 140 & 30 & & \\
      & 2 & 1 & 180 & 48 & & \\
      & 2 & 3 & 140 & 22 & & \\
\hline
$2^-$ & 2 & 1 & 180 & 59 & & \\
      & 2 & 3 & 140 & 39 & & \\
      & 4 & 5 & 80  & 1 & & \\
\hline
$3^-$ & 0 & 3 & 140 & 28 & & \\
      & 2 & 1 & 180 & 65 & & \\
      & 2 & 5 & 100 & 1 & & \\
      & 4 & 1 & 80  & 5 & & \\
\hline
$4^-$ & 2 & 3 & 160 & 72 & & \\
      & 2 & 5 & 160 & 1 & & \\
      & 4 & 1 & 160 & 26 & & \\
\hline
\end{tabular}
\end{center}
\end{footnotesize}
\end{table}

Convergence has been achieved with the sets of quantum numbers shown
in table \ref{tab1}.  The large values of $K_{max}$ should roughly be
divided by two to give the corresponding number of basis states in
each partial wave.  This means that the total number of basis states
is about $3 \times 90$ for $\ell_x=\ell_y=0$, and in total for the
$0^+$-resonances the basis consists of about 700 states. Some of the
other resonances have a larger number of basis states.  With present
day computers these numbers are far from being large but still
extremely efficient compared to the use of only one Jacobi system and
all possible partial waves consistent with one value of $K_{max}$
\cite{fed97}.  These high partial waves are needed for one Jacobi
system to describe configurations with two spatially close-lying
particles relatively far from the third particle.  The Faddeev
decomposition takes care of that in the present formulation.

A given angular momentum and parity of an intended $^{12}$C-state
needs a set of these partial wave components.  However, more than one
$^{12}$C-state with this spin and parity may be found from such a set.
The probability to find each partial wave for each resonance is also
given in table \ref{tab1}. We shall postpone the discussion of these
values until section 3.4.

\subsection{Adiabatic potentials}

The adiabatic potentials result from a full quantum mechanical
solutions of the angular part of the Faddeev equation for fixed
hyperradius.  Each potential corresponds to a specific combination of
the partial waves, in some cases strongly varying with $\rho$. These
structures form the basis for the radial solution which truly is a
resonance in the three-body continuum.  Very few angular eigenvalues
are usually needed for convergence, and their behavior are decisive
for the resulting resonances.  This is not a trivial conclusion, since
each point of these potentials by definition corresponds to the same
$\rho$, but different configurations are otherwise allowed. It is not
obvious at all that the structures of the resonances predominantly
arise from these combinations of configurations contributing to each
adiabatic potential.

\begin{figure*}[h]
\epsfig{file=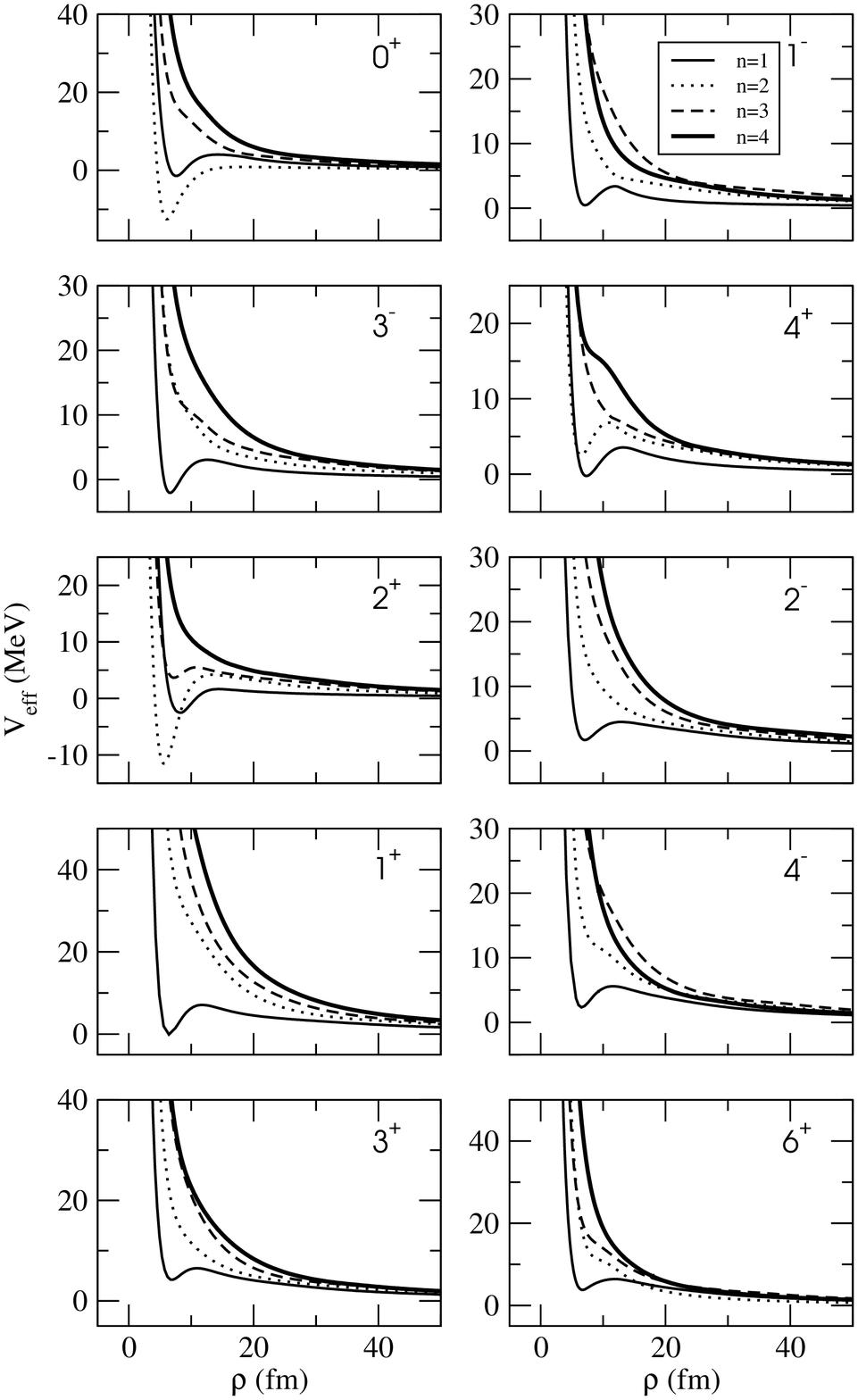,scale=0.75,angle=0}
\caption{The real parts of the four lowest adiabatic effective 
potentials, including the three-body potentials, as functions of $\rho$
for the $^{12}$C resonances with $J^{\pi}$ given in the figures. The
two-body interaction, obtained from \cite{fed96}, is a slightly
modified version of the a1-potential of \cite{ali66}.  The parameters
($S$ and $b$) of the three-body Gaussian potentials, $S
\exp(-\rho^2/b^2)$, are given in table
\ref{tab2}. }
\label{potentials}
\end{figure*}

We show in fig. \ref{potentials} the real parts of these potentials
including $V_{3b}$ corresponding to the quantum numbers of each of the
resonances.  The imaginary parts are small, oscillate around zero and
vanish at large $\rho$.  All these potentials diverge as $\rho^{-2}$
when $\rho
\rightarrow 0$ due to the generalized centrifugal barrier. They vanish 
as $1/\rho$ for large $\rho$ where Coulomb is the only contributing
interaction.  At relatively small distances the potentials have minima
supporting the bound states and resonances.  The sizes of these
pockets are larger than obtained from the two-body interaction which
in most cases is far from sufficient to place the resonance at the
correct energy.  Therefore we added in each case a diagonal,
short-range three-body potential designed to mock up effects of the
intrinsic degrees of freedom in a $3\alpha$-model at short distances
where the three $\alpha$-particles overlap.  This allows us to stay
within the three-body model and still provide the proper boundary
condition at small distance at the right energy.

The three-body potentials are individually adjusted to reproduce one
of the energies, but chosen to be the same for all adiabatic
potentials of given angular momentum and parity.  We typically choose
the best known of the resonances since it is the continuum state of
lowest energy, i.e. closest to the threshold for 3 free
$\alpha$-particles.  Obviously the bound states are better known but
their structures can easily differ even more from the
$3\alpha$-structure than the resonance states. The three-body
potentials could then differ substantially and predictions for other
resonances probably would be less accurate. The systems with
wavefunctions localized outside the overlap region for the
$\alpha$-particles should need less three-body potential and then be
more reliably calculated in the $3\alpha$-model.  For example the
ground state could be a spatially confined structure while
$0^+$-resonances could be fairly good three-body states.

For the $0^+$, $2^+$ and $4^+$-cases two or three minima appear in
different potentials indicating the possibility of different
structures of low-lying resonances.  For the $1^+$-case the minimum
only appears after addition of the three-body potential indicating
that the corresponding structure at small distance differs
substantially from three $\alpha$-particles.  All negative parity
states exhibit one minimum at distances larger than the $6$~fm chosen
as the range of the three-body interaction.

\subsection{Resonance energies}

The length coordinate, $\rho$, is rotated in the complex plane and the
adiabatic potentials including non-diagonal terms are computed.  By
adding the diagonal three-body potential all quantities are specified
in the coupled set of radial equations.  The solutions are the
three-body resonances where key quantities are real and imaginary
parts of the energies. The strength of the three-body potential is
used to move one bound state or one resonance into a preferred
position for each set of angular momentum and parity.  If there is
more than one solution they are moved simultaneously by the same
potential.  The range of the three-body potential is kept at the same
value implying that the width is predicted and not fitted.  The
associated uncertainties are discussed in section 4.

\begin{figure*}[h]
\epsfig{file=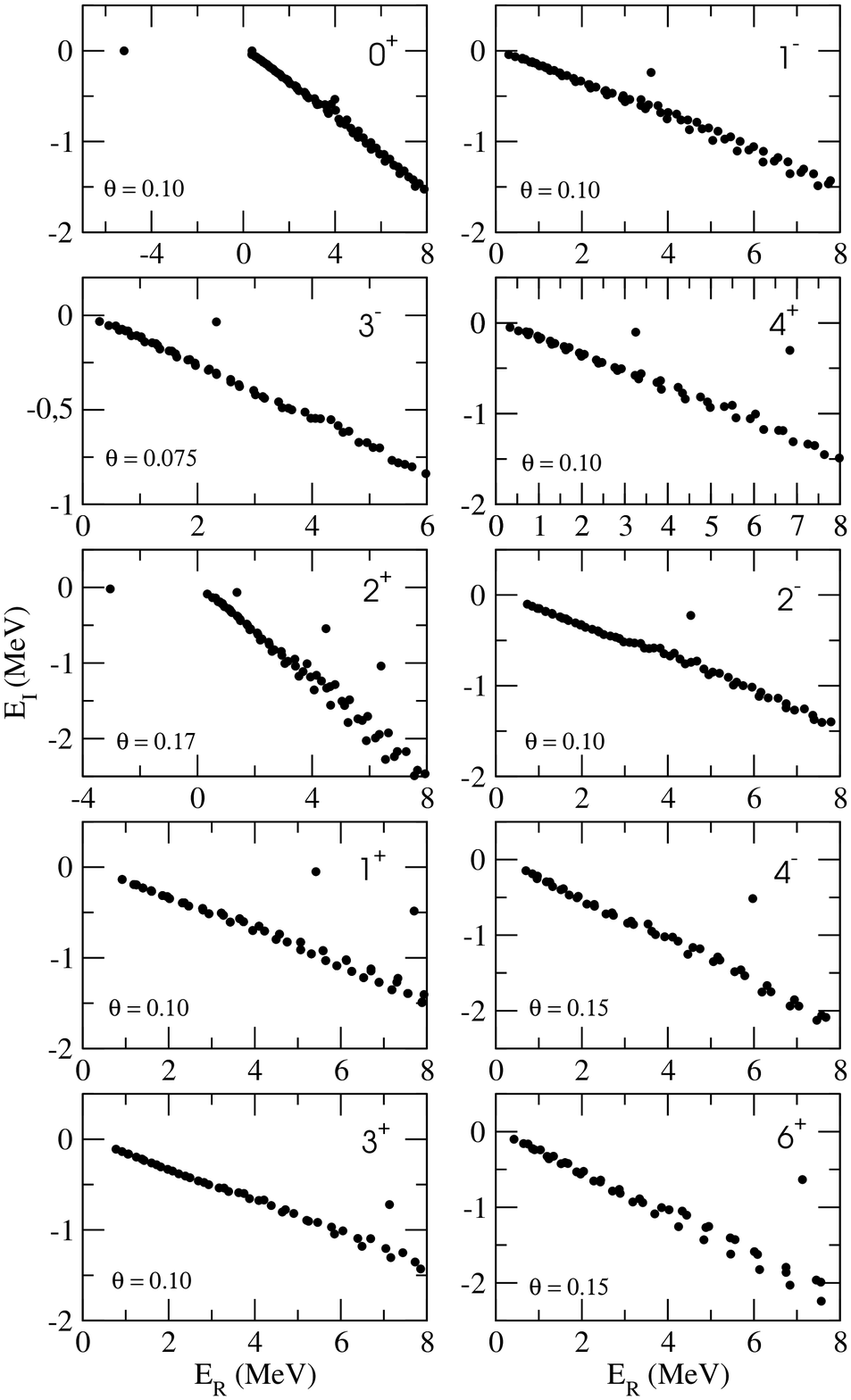,scale=0.75,angle=0}
\caption{The real ($E_R$) and imaginary ($E_I=-\Gamma/2$) parts of the 
energies for $J^{\pi}$-resonances in $^{12}$C obtained after complex
rotation by the angle $\theta$ also given in the figure. The two and
three body interactions are specified in fig. \ref{potentials}.  }
\label{spectr}
\end{figure*}

The computed results for the resonance energies are shown in
fig. \ref{spectr}. The individual figures all exhibit the same
dominant features with a series of points forming almost a straight
line.  Each of these points corresponds to a continuum state.  These
discretized three-body continuum states appear due to an imposed
condition of zero radial wavefunction at the boundary of a large
box. The energies appear on the straight line arising by rotation of
an angle $2 \theta$ around the origin.  The fluctuations are due to
numerical uncertainties.

\begin{table}[t]
\caption{Calculated and measured energies $E_R$ (in MeV) and partial 
$\alpha$-decay widths $\Gamma_R$ (in keV) of the resonances for 
different $J^\pi$. Since our formalism cannot
describe an isospin one state, we found only one $1^+$ resonance whose energy
has been adjusted to reproduce both measured values. Experimental values
(labeled ``exp'') are from \cite{ajz90,dig05,dig06} and calculated
results (labeled ``th'') are obtained with the two-body interaction
``AB(a')'' and the three-body interaction parameters $S$ (in MeV) and
$b$ (in fm) given here. The energies are measured from the $3\alpha$
threshold.  The computations are for a rotation angle $\theta$ (in
rad).  The $1^+$ state at 7.84 MeV labeled $T=1$ has isospin one. }
\label{tab2}
\vspace*{0.2cm}
\begin{tabular}{|cccccccc|}
\hline
$J^\pi$&$E_{R,exp}$&$\Gamma_{R,exp}$&$E_{R,th}$&$\Gamma_{R,th}$&S&b&$\theta$\cr
\hline
$0^+$ & -7.25 & 0.0 & -5.19 & 0.0 & -20.0 & 6& 0.1\\
      & 0.38  &  0.0085  & 0.38 &  0.0625  &-20.0 &6&0.1\\
      &   4.3  &  3490  & 3.95 & 1000 & -20.0 & 6 & 0.1 \\
\hline
$1^-$ & 3.57 & 315 & 3.61 & 475 &-2.8&6&0.1\\
\hline
$3^-$ & 2.37 & 34 & 2.33 & 68 &-1.7 & 6&0.075\\
\hline
$2^+$ &-2.875 & 0.0 & -3.04 & 0.0 &-17.0 &6&0.17\\
      &  &  & 1.38 & 132 &-17.0&6&0.17\\
      & 3.88     & 430     & 4.48 & 1086 &-17.0&6&0.17\\
      &  6.3 & 1700 & 6.49 & 2250 &-17.0&6&0.17\\
\hline
$4^+$ &      &     & 3.25 & 396 &-25.5&6&0.1\\
& 6.81 & 258 & 6.83 & 606 &-25.5 & 6&0.1\\
\hline
$6^+$ &  &  & 7.13 & 1267 & -20 & 6 & 0.15 \\
\hline
$2^-$ & 4.55 & 260 & 4.53 & 452 &-2.8&6&0.1\\
\hline
$1^+$ & 5.43 & 0.0177 & 5.42 & 48.6 &-92&6&0.1\\
\tiny{T=1}& 7.84 & 0.0018 & 7.70 & 948 &-78 &6&0.1\\
\hline
$3^+$ &  &  & 7.13 & 1450 & -20  & 6  & 0.1 \\
\hline
$4^-$ & 6.08 & 375 & 5.98 & 1035 &-1.8&6&0.15\\
\hline
\end{tabular}
\end{table}

The points above the straight line are the resonance energies
appearing with positive real parts, $E_R$, and negative imaginary
parts, $E_I = - \Gamma/2$, where $\Gamma$ is the width of the
resonance.  The radial wavefunctions of these resonances vanish
exponentially at large distances corresponding to outgoing waves in all
channels.  This boundary condition is identical to those valid for the
bound states which appear on the negative side of the real axis.

The computed resonance energies are collected in table \ref{tab2}
where the experimental values also are given.  Three $0^+$-states
appear, i.e. the ground state and two resonances.  The first resonance
is very close to the threshold with an extremely small width of about
62~eV.  It is then difficult to distinguish from a discretized
continuum state without information from its wavefunction.  The second
$0^+$-resonance appears around 4~MeV with a width of about 1~MeV. The
real energy disagrees with previous experiments \cite{ajz90}, but it
is consistent with the analysis of the recent experiment
\cite{dig05,dig06} as quoted in table \ref{tab2}.

We find four $2^+$-states, i.e. one excited bound state and three
resonances. Only two resonances are suggested by experiments
\cite{ajz90,dig05,dig06}.  With the natural choice of $b=6$~fm the
highest of the computed resonances can easily be adjusted to the
experimental value \cite{dig05,dig06}.  It would perhaps be natural to
adjust the lowest resonance to the measured energy.  However, an
attempt to do this failed because a smaller attraction than given in
table \ref{tab2} move the energy upwards but at the same time the top
of the confining barrier is approached and exceeded leading to much
faster increase of the width.  The result is that the width increases
quickly whereas the energy remains essentially unchanged.  The same
behavior is seen for the second of the $2^+$-resonances.  Only the
third follows naturally the experimental specifications with this
choice of one three-body interaction of the present Gaussian form.

We find two $4^+$-resonances. If the three-body strength is reduced
from the value in table \ref{tab2} the widths of both resonances
increase dramatically around 4~MeV and 9~MeV, respectively.  Therefore
the most consistent solution is that the highest $4^+$-resonance is
found experimentally while the lowest has been hidden behind other
states in previous experiments.  Consequences of allowing variation of
$b$ are discussed in section 4.

Two $1^+$ resonances are known experimentally, where the 7.84 MeV state is
an isospin $T=1$ state. The two calculated $1^+$-resonances are
artificial in the sense that only one of these resonances is found in
the calculation, since a state with $T=1$ cannot be described by our
formalism. The energy is then adjusted through the three-body
potential to either of the measured values corresponding to isospin
zero or one.  The fact that only one state is found in the computation
reflects that we only have the correct degrees of freedom to describe
$T=0$ states. The reason for also adjusting to the $T=1$ energy is the
expectation that the decay into three $\alpha$-particles would occur
through the same intermediate and large-distance structures as for
$T=0$.  We do not imply that the $T=1$ state has the same structure at
small distances.

All other states than $J^{\pi}=0^{+},2^{+},4^{+}$ only appear once.
Their energies are then adjusted by use of the three-body potential
and the related width for the range of $b=6$~fm can be compared to the
measured values. Since only one $2^-$ and one $4^-$-resonance appear
we suggest that the uncertain assignment of the second experimental
$2^-$-resonance should be changed to $4^-$. The $3^{+}$ and $6^{+}$
states are not known experimentally and we can only make rough guesses
of the three-body strength.  Since $6^{+}$ belongs to the family of
even angular momenta and even parity we used a similar strength of
$-20$~MeV.  The unnatural parity of $3^{+}$ puts it in the same class
as the $1^{+}$ state but the strength of $-20$~MeV already leads to an
energy of about 6~MeV. This $3^{+}$ estimate is then very uncertain.

It is remarkable that the ratio between theoretical and experimental
values for the widths in almost all cases is around 2.  The
widths are in most cases fairly well estimated by the tunneling
probabilities through the respective barriers shown in
fig.~\ref{potentials}, see \cite{fed03a}. This corresponds to an
exponential dependence on energy and deviations within a factor of two
can then be considered to be very accurate \cite{fed03a}. These
results are obtained without special treatment of the small distance
properties beyond that corresponding to the $3\alpha$-structures.  The
possibility of preformation factors or $\alpha$-particle spectroscopic
factors, as appearing in ordinary $\alpha$-decay, seems to play a
minor role. It is interesting to note that we systematically calculate
slightly too large widths indicating preformation factors at the
surface of about $1/2$.

Two exceptions from the general rule are seen in the second $0^+$ and
the two $1^+$ resonances, where the computed width is 3.4 times
smaller for $0^+$ and larger for the two $1^+$ states by $2.6 \cdot
10^{3}$ and $5.3 \cdot 10^{5}$, respectively. The $1^+$ states require
in contrast to all the other states preformation (spectroscopic)
factors three and five orders of magnitude smaller than for the other
states.  This reflects totally different structures corresponding to a
full break down of the $3\alpha$-descriptions at small distances.
This is particularly pronounced for the $T=1$ state which cannot be
described by the $T=0$ $\alpha$-particle building blocks. The
preformation factor correspondingly comes out as exceedingly small.

The different structure is less obvious for the lowest $1^+$ state
with $T=0$ which is not forbidden for symmetry reasons in the $3
\alpha$-models.  However, a strong hint is found in the model by the
unusually large lowest possible hypermomentum of $8$. The immediate
guess of $K=1$ for $1^+$ states are ruled out, together with all other
hypermoments below $K=8$, by the imposed boson symmetry of the three
$\alpha$-particles \cite{kor90}. This is in clear contrast to all the
other $J^{\pi}$ states where the lowest hypermomentum consistent with
angular momentum always contribute, see table \ref{tab1}.

These results are also, apart form the $1^+$ anomalies, remarkable for
at least one other reason, i.e.  a monotonic dependence on energy,
angular momentum and parity does not exist.  The barriers in
hyperradius provided by the calculation must then be physically
reasonable, in turn implying that the concept of the hyperspheric
adiabatic expansion is in rather close agreement with the physical
process.  The systematic correlation with the experimental widths,
varying by orders of magnitude, is otherwise totally unexplained.  It
should here be emphasized that many of the resonances are not well
described as cluster states \cite{kan98}.  This is particularly clear
for the two $1^+$ states.

The three-body potentials with $b=6$~fm seem to fall in two groups,
i.e. one with a strength $S$ of about $-20$~MeV ($0^+$, $2^+$, $3^+$,
$4^+$, $6^+$) and one with $S \approx -2$~MeV ($1^-$, $2^-$, $3^-$,
$4^-$).  This should indicate that the amount of $\alpha$-cluster
states within each group is roughly the same as also reflected by the
comparison of calculated and known widths.  Exceptions are the $1^+$
states with much larger strengths of about $-80$~MeV. The computed
widths for these states increase substantially with the range $b$ of
the three-body parameter when the strengths correspondingly are
increased to maintain the energy. The apparent spectroscopic factor
for the isospin zero state would for $b \approx 4$~fm be comparable to
the values for the other states as reported in \cite{fed03a}.

\begin{table*}[t]
\caption{\label{tab3}
Sizes expressed as root mean square radii (in fm) of the $^{12}$C
bound states and resonances.  The second column is $\rho_{rms} =
|<\rho^2>|^{1/2}$, and the fourth column is without absolute squares
of the expectation values.  The third and fifth columns are root mean
square radii of $^{12}$C states obtained by eq.~(\ref{e137}) with the
two different $<\rho^2>$ prescriptions, respectively.  The last four
columns are computations from \cite{kan98,kam81,ueg79,fun05}.  The
root mean square charge radius of $^{12}$C ground state is
$<r^2>^{1/2} = 2.4829 \pm 0.0019$~fm
\cite{ruc84}.  }
\begin{center}
\begin{tabular}{|l|cc|cc|cccc|}
\hline
$J^\pi$& $\rho_{rms}$ & $r_{rms}$& $<\rho^2>^{1/2}$& $r_{rms}$ &
$r_{kan}$ \cite{kan98} & $r_{kam}$ \cite{kam81} & $r_{ueg}$ \cite{ueg79} 
& $r_{fun}$ \cite{fun05} \\
\hline
$0^+_0$ &7.39&2.59 &  7.35-i0.74&2.59-i0.00 & 2.53 & 2.468 & 2.40 & 2.40 \\
$0^+_1$ &12.19&3.81 & 12.14-i1.16& 3.82-i0.02 & 3.27 & 3.56 & 3.40 & 3.83 \\
$0^+_2$ &11.35&3.59 & 10.80+i3.50& 3.30+i1.20 & 3.98 & & 3.52 &\\
\hline
$1^+$ &8.41&2.84 & 8.40-i0.43& 2.84+i0.10 & 2.47 & & &  \\
\hline
$3^+$ & 8.95 & 2.97 & 8.79+i1.70 & 2.86+i0.64 & & & &\\
\hline
$2^+_0$ &6.79&2.45 & 6.69-i1.14& 2.45+i0.00 & 2.66 & 2.51 &2.36 & 2.38  \\
$2^+_1$ &11.37&3.60 & 11.37+i0.11& 3.54+i0.53 & 3.99 & & 3.52 &   \\
$2^+_2$ &11.90&3.74 & 11.73+i2.00& 3.53+i1.05 & 3.50 & 4.1 & 3.34 & \\
$2^+_3$ &11.90&3.74 & 11.73+i2.00& 3.53+i1.05 & 3.86 & & &  \\
\hline
$4^+_1$ & 10.19 & 3.29 & 10.19+i0.11& 3.27+i0.29&  2.71 &  2.31 & 2.29 & 2.31\\
$4^+_2$ & 8.68 & 2.91 & 8.68-i0.01 & 2.89+i0.21& 4.16 & 2.44 & 3.64 & \\
\hline 
$6^+$ & 9.49 & 3.11 & 9.44+i1.04 & 3.01+i0.62 & & & & \\
\hline
$1^-$ &10.24&3.30 & 10.22+i0.66& 3.26+i0.43 & 3.42 & 3.45 & 3.29 & \\
\hline
$2^-$ &10.11&3.27 & 10.02+i1.39& 3.18+i0.61 & 3.49 & & 3.32 & \\
\hline
$3^-$ &9.33&3.07 & 9.33+i0.10& 3.06+i0.20 & 3.13 & 2.87 & 2.83 & \\
\hline
$4^-$ &9.03&2.99 & 9.02+i0.51& 2.93+i0.47 & 3.19 & & 2.87 &\\
\hline
\end{tabular}
\end{center}
\end{table*}

Beside the energy the size is a characteristic quantity for any
system. This is straightforward for bound states but resonances decay
and are therefore infinitely large. However, the complex scaling
provides a simple measure of the extension of the small distance part
of the wavefunction, i.e. expectation values related to the
hyperradius.  The usual relation is \cite{nie01}
\begin{eqnarray}  
  r_{rms}^{2} =    \frac{1}{12} \langle \Psi | \rho^{2} | \Psi \rangle 
 \exp(i 2\theta) +   R_{\alpha}^{2}   \label{e137} \;, 
\end{eqnarray} 
where $R_{\alpha}=1.47$~fm and $r_{rms}$ are the root mean square
radii of the $\alpha$-particle and $^{12}$C in the $3\alpha$-model,
respectively.  The finite size of the three $\alpha$-particles are
then accounted for.  Complex rotation by $\theta$ means that the
square of the hyperradius is multiplied by $\exp(i 2\theta)$.  The
expectation value in eq.(\ref{e137}) is then expressed in the rotated
system, and now this quantity is finite.  They are still complex
numbers and the absolute square can be used as the measure of
size. Another tempting possibility is to interpret the real part as
the radius of the resonance and the imaginary part as the variation
(uncertainty) of the radius of the continuum wavefunctions as energy
changes through the interval $E_R \pm \Gamma/2$ around the central
energy \cite{moy98}.

The numerical results for the resonances and bound states are given in
table \ref{tab3} for both prescriptions.  The real parts are in all
cases very close to each other and certainly within the
``uncertainty'' measured by the imaginary part.  For the two bound
states the imaginary part reflects the small uncertainty in the
numerical computation, since here the values should strictly be real.
The ground state result is within the uncertainty identical to the
root mean square radius of $^{12}$C, i.e. 2.468~fm. The
``uncertainties'' are small for narrow resonances and relatively large
for broad resonances.  The bound $2^+$-state is smaller than the
ground state and all the resonances are larger. The $0^+$ and
$2^+$-resonances are especially large because they are built of the
same partial waves and orthogonality then apparently push them
outwards. The comparison to the old results derived from the
resonating group method \cite{kam81} and the ones from the generator coordinate
calculation \cite{ueg79}, as well as the recent computation
\cite{fun05} and the
antisymmetrized molecular dynamics results from
\cite{kan98} are in agreement within the uncertainties.  It should be 
emphasized that a detailed comparison should take the energy
differences into account.  From halo physics of neutral particles we
know that energy and radius are closely correlated.  The radial
variation with binding energy is much smaller for charged particles
and finite angular momentum.

\subsection{Resonance structures}

\begin{figure*}[h]
\epsfig{file=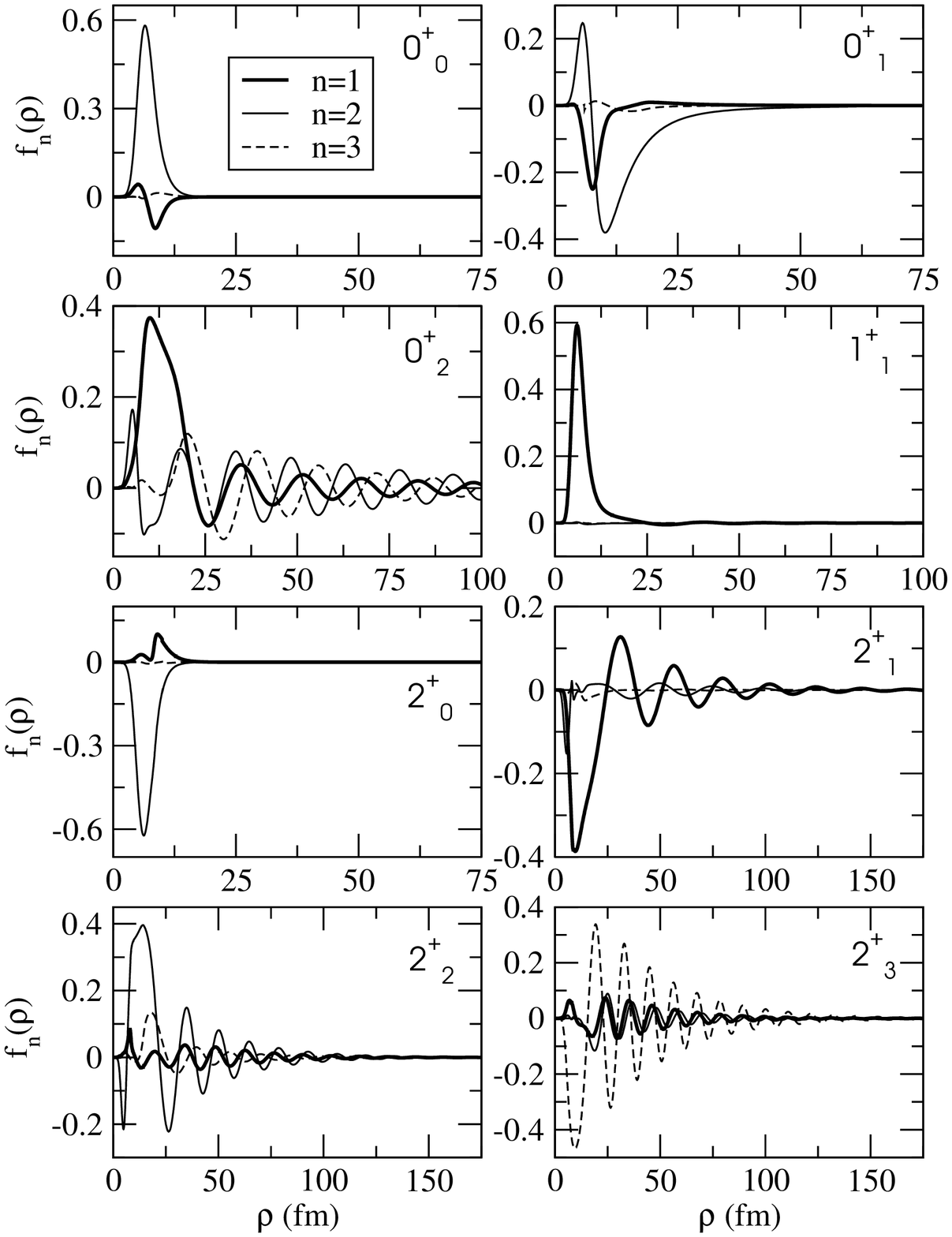,scale=0.75,angle=0}
\caption{The real parts of the three lowest adiabatic 
radial wavefunctions as functions of $\rho$ for the $0^{+}$, $1^{+}$
and $2^{+}$-states of $^{12}$C.  Their normalization reflects their
relative contribution to each resonance.  The two and three-body
interactions are specified in fig. \ref{potentials}.}
\label{rada}
\end{figure*}

\begin{figure*}[h]
\epsfig{file=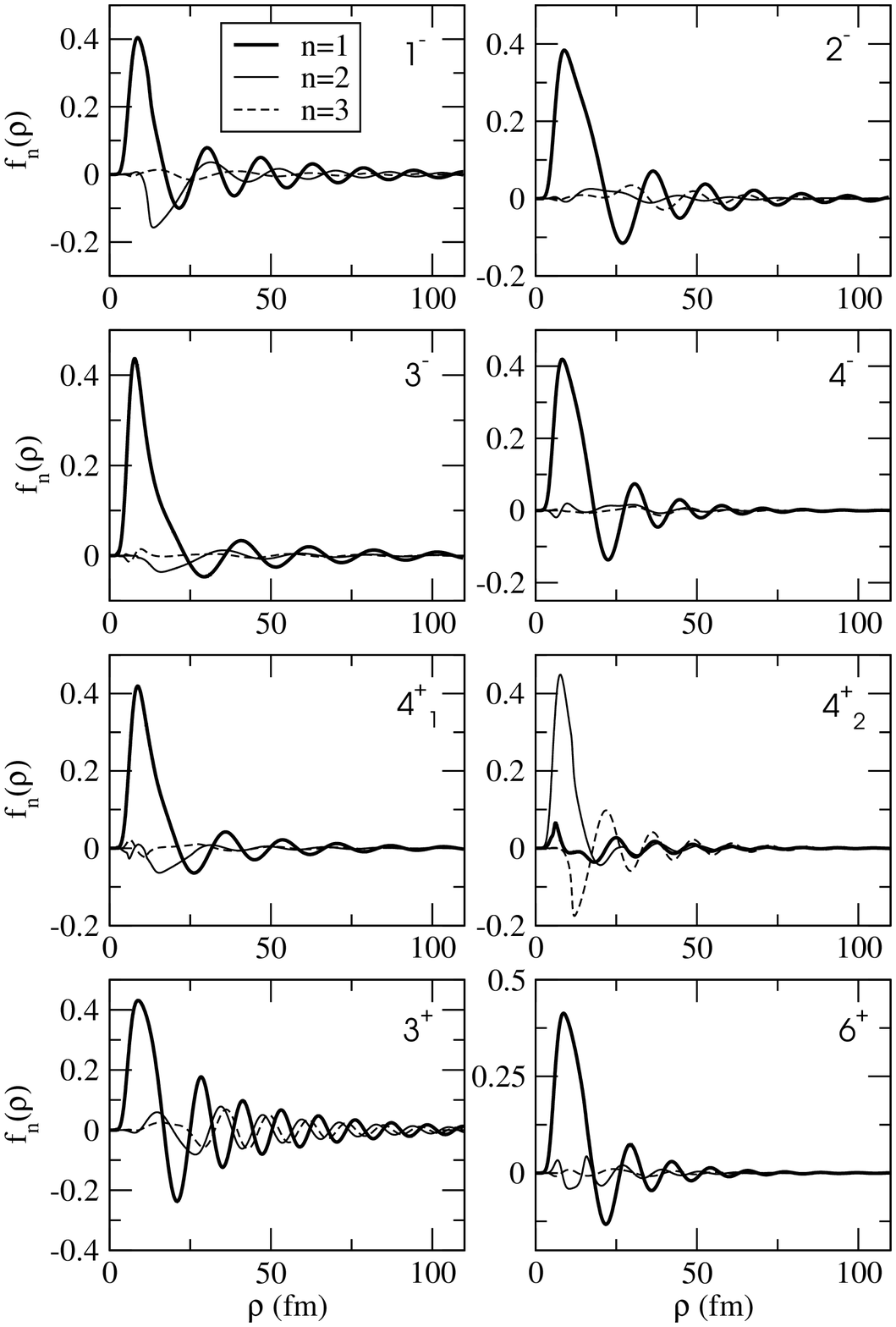,scale=0.75,angle=0}
\caption{The same as fig. \ref{rada} but for the $1^{-}$, $2^{-}$, 
$3^{-}$, $4^{-}$, $4^{+}$, $3^{+}$ and $6^{+}$-states of $^{12}$C. }
\label{radb}
\end{figure*}

The resonance structure is determined by combining the amplitude
obtained as the solution to the radial wavefunction with the
individual structure of each adiabatic component.  The real parts of
the radial wavefunctions are shown in figs. \ref{rada} and \ref{radb}
for the two bound states and each of the 14 resonances.  At least two
general features are striking. First, only very few of these
wavefunctions contribute significantly. Furthermore, for all states
one adiabatic component is much larger than all other contributions.
This reflects the rather fast convergence of the adiabatic expansion
where three eigenvalues often provide an accurate solution. Second,
the wavefunctions oscillate with exponentially decreasing amplitudes
as functions of $\rho$.  This behavior is qualitatively the same for
bound states and resonances which precisely is the reason for using
complex scaling to compute resonance wavefunctions and energies.  The
imaginary parts are not shown to avoid cluttering the figures. They
exhibit similar oscillatory behavior.

\begin{figure*}
\epsfig{file=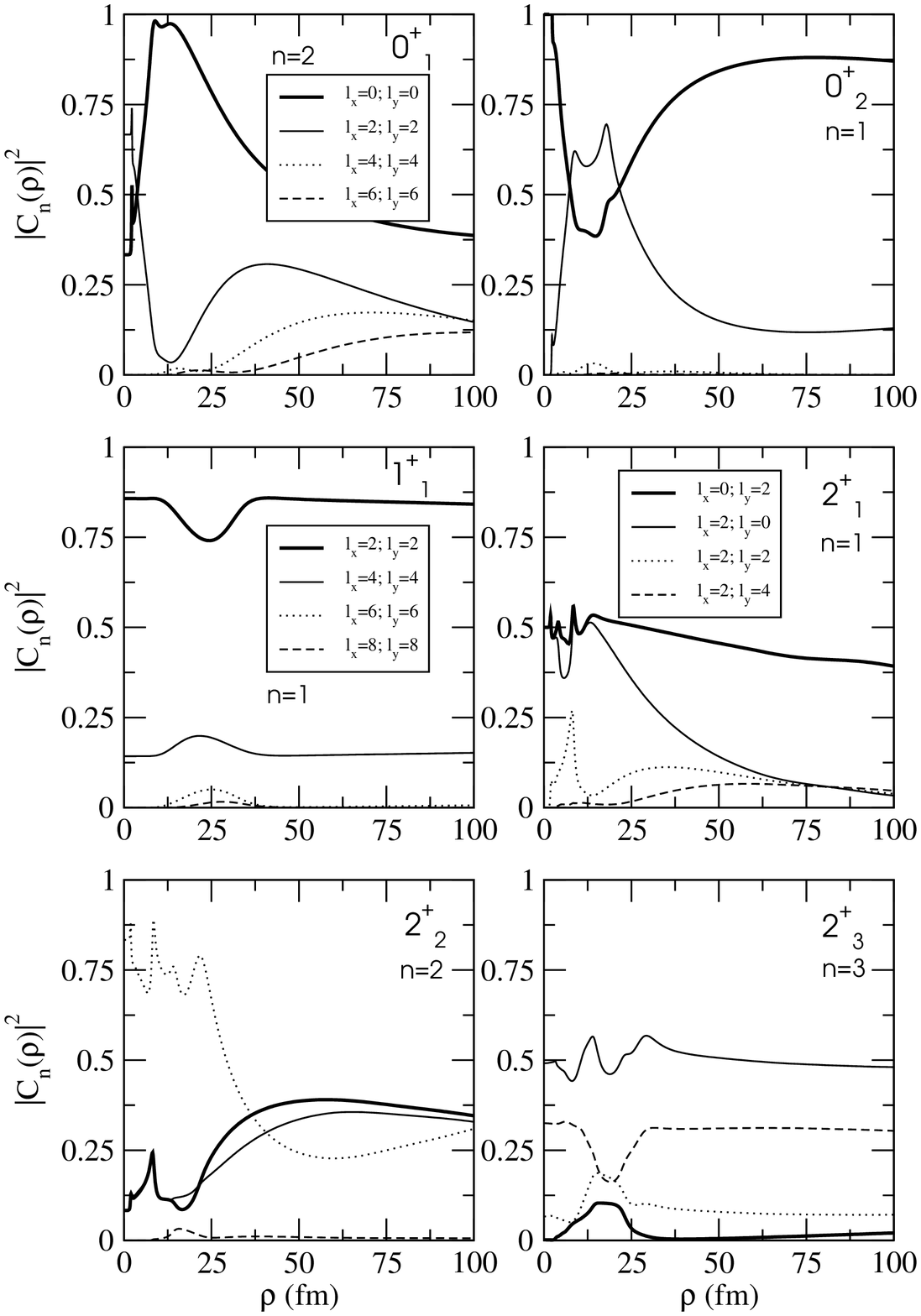,scale=0.75,angle=0}
\caption{The partial wave decomposition of the $^{12}$C 
resonances with $J^{\pi}$ as indicated on the figure shown as function
of $\rho$ for the dominating adiabatic eigenvalue. The two and three-body
interactions are specified in fig. \ref{potentials}. }
\label{contra}
\end{figure*}

\begin{figure*}
\epsfig{file=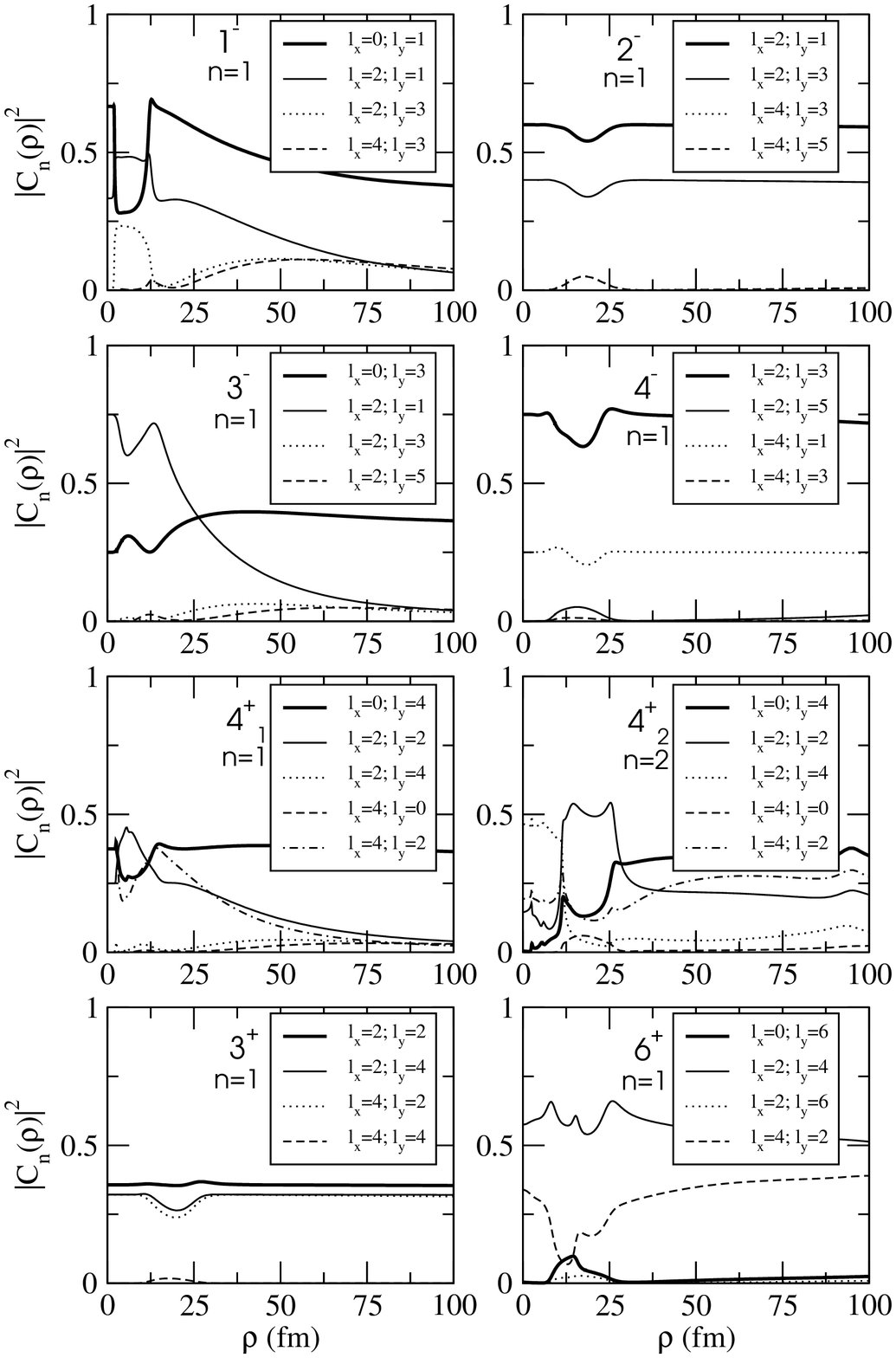,scale=0.75,angle=0}
\caption{The partial wave decomposition of the $^{12}$C 
resonances with $J^{\pi}$ as indicated on the figure shown as function
of $\rho$ for the dominating adiabatic eigenvalue. The two and
three-body interactions are specified in fig. \ref{potentials}.  }
\label{contrb}
\end{figure*}

The probabilities for finding the different adiabatic components in
each resonance wavefunction are given in table \ref{tab4}.  The
structures of the resonances are then essentially contained in these
dominating angular wavefunctions, where each is related to one
adiabatic potential.  For each resonance we therefore take the
dominating contribution and decompose into the partial waves specified
in table \ref{tab1}. These partial wave decompositions are shown in
figs. \ref{contra} and \ref{contrb}.  The overall feature is that the
dependence on hyperradius is rather strong. However, we should
emphasize that the small distance structure only has little physical
significance.  Numerically the variation arises from couplings of
low-lying adiabatic potentials which change rapidly in this region due
to the different behavior of $s,d$ and $g$ two-body potentials imposed
by the different behavior of the phase shifts.

The two dominating wavefunctions for the $0^+$-resonances have very
different behavior.  The first is dominated by $\ell_x=\ell_y=0$ at
about 10~fm, but at larger $\rho$ the higher partial waves all become
almost equally important.  The large-distance behavior is consistent
with the description of one $\alpha$-particle far away from a
spatially confined $^{8}$Be($0^+$)-structure.  The higher partial
waves arise from $s$-waves in the ``unnatural'' Jacobi systems.  The
second wavefunction has about equal contributions at small $\rho$ from
$\ell_x=\ell_y=0,2$, but at large $\rho$ the $s$-wave component
increases strongly while the higher partial waves correspondingly
decrease towards zero.  This large-distance structure is consistent
with all three $\alpha$-particles symmetrically distributed far from
each other.

\begin{table}
\caption{\label{tab4}
Contribution $W_n$ of each adiabatic potential, labeled by $n$, to the total
norm of the wave function of each $^{12}$C-state.  Only those
contributions of more than 1\% are given.  }
\begin{center}
\begin{tabular}{|cccc|}
\hline
$J^\pi$& $W_{1}$ & $W_{2}$& $W_{3}$  \\
\hline
$0^+_0 $ & 2 & 98 &   \\
$0^+_1$& 15  &84  &  \\
$0^+_2$& 96  &4  &  \\
\hline
$1^+$& 100 &  &  \\
\hline
$2^+_0$ &  & 100   & \\
$2^+_1$&100  &  &  \\
$2^+_2$&  & 94  &7  \\
$2^+_3$&1  &  &99  \\
\hline
$3^+$& 99 & 1 &   \\
\hline
$4^+_1$& 97 & 3  &\\
$4^+_2$&  & 97  & 3  \\
\hline
$6^+$&  &  &  \\
\hline
$1^-$& 88 &12  &  \\
\hline
$2^-$&99  &  &  \\
\hline
$3^-$&99  &1  &  \\
\hline
$4^-$&100  &  &  \\
\hline
\end{tabular}
\end{center}
\end{table}

The first of the other natural parity wavefunctions, corresponding to
$1^-$, $3^-$, $2^+$, $4^+$ and $6^+$, all have $\ell_x=0$ as the
dominating component at large distance.  This is consistent with a
$^{8}$Be($0^+$)-structure and the third particle with $\ell_y$ equal
to the total angular momentum.  Several other components are small and
of comparable magnitude. The lowest $0^+$-wavefunction also falls into
this category.

The first wavefunctions for unnatural parity, $1^+$, $3^+$, $2^-$,
$4^-$, all have $\ell_x=2$ as the dominating component both at small
and large distances.  The angular momentum, $\ell_y$, of the third
particle for this component is as small as possible consistent with
the total spin and parity.  Only one more component gives significant
contributions for each of these states, i.e. $\ell_x=4$ combined with
the lowest possible $\ell_y$ for $1^+$ and $4^-$, and $\ell_x=2$ and
$\ell_y=3$ for $2^-$. The latter component is larger than the
contribution from the higher partial wave of $\ell_x=4$ and the lowest
possible $\ell_y$-value of $3$.

The third $2^+$-wavefunction is similar in structure with one
dominating component of $\ell_x=2$ and $\ell_y=0$, and another
component of half the size with $\ell_x=2$ and $\ell_y=4$.  Here the
component with $\ell_x=2$ and $\ell_y=2$ only contributes
insignificantly.  The second $2^+$-wavefunction has three about equal
contributions at large distance from $(\ell_x,\ell_y) =
(0,2),(2,0),(2,2)$.  The second $4^+$-wavefunction also has three
comparable contributions from $(\ell_x,\ell_y) = (0,4),(4,0),(2,2)$,
and in addition several smaller components.

We emphasize that the contributions from the individual adiabatic
potentials should be combined with their respective amplitudes to give
the total structure of the resonances.  These amplitudes depend on
hyperradius and the probabilities in table \ref{tab4} are averages
over all $\rho$. The exponential decrease with $\rho$ then obviously
heavily enhances the structures at the small distances.  The relative
distributions at larger distances may change substantially.

The different partial waves receive contributions from different
adiabatic components. The probabilities given in table \ref{tab2} can
be obtained by combining the information in figs. \ref{contra} and
\ref{contrb}, and figs. \ref{rada} and \ref{radb}. The results reflect 
the conclusions from the above discussion.

\section{Interaction dependence}

We have to distinguish between two types of inaccuracies, i.e. arising
from (i) inaccuracies in the numerical computations for given sets of
input parameters, and (ii) model assumptions and uncertainties in the
corresponding parameters.  Here the uncertainties in (i) originate
from cut-offs in different basis expansions, i.e. the partial waves
and the number of Jacobi polynomials for each of the three Faddeev
components, and the number of adiabatic potentials.  The convergence
can be seen directly in the decreasing amplitudes for both partial
waves and adiabatic potentials. We emphasize that the contributing
terms vary with hyperradius as discussed in the previous section.
Also the rotation angle should be chosen as small as possible to
optimize accuracy while still allowing distinction of the three-body
resonance from the background continuum states.

Model uncertainties are more difficult to assess. We shall first
consider those intrinsic to the cluster model, and afterwords compare
to experiments and other more microscopic computations.

\begin{figure*}[h]
\epsfig{file=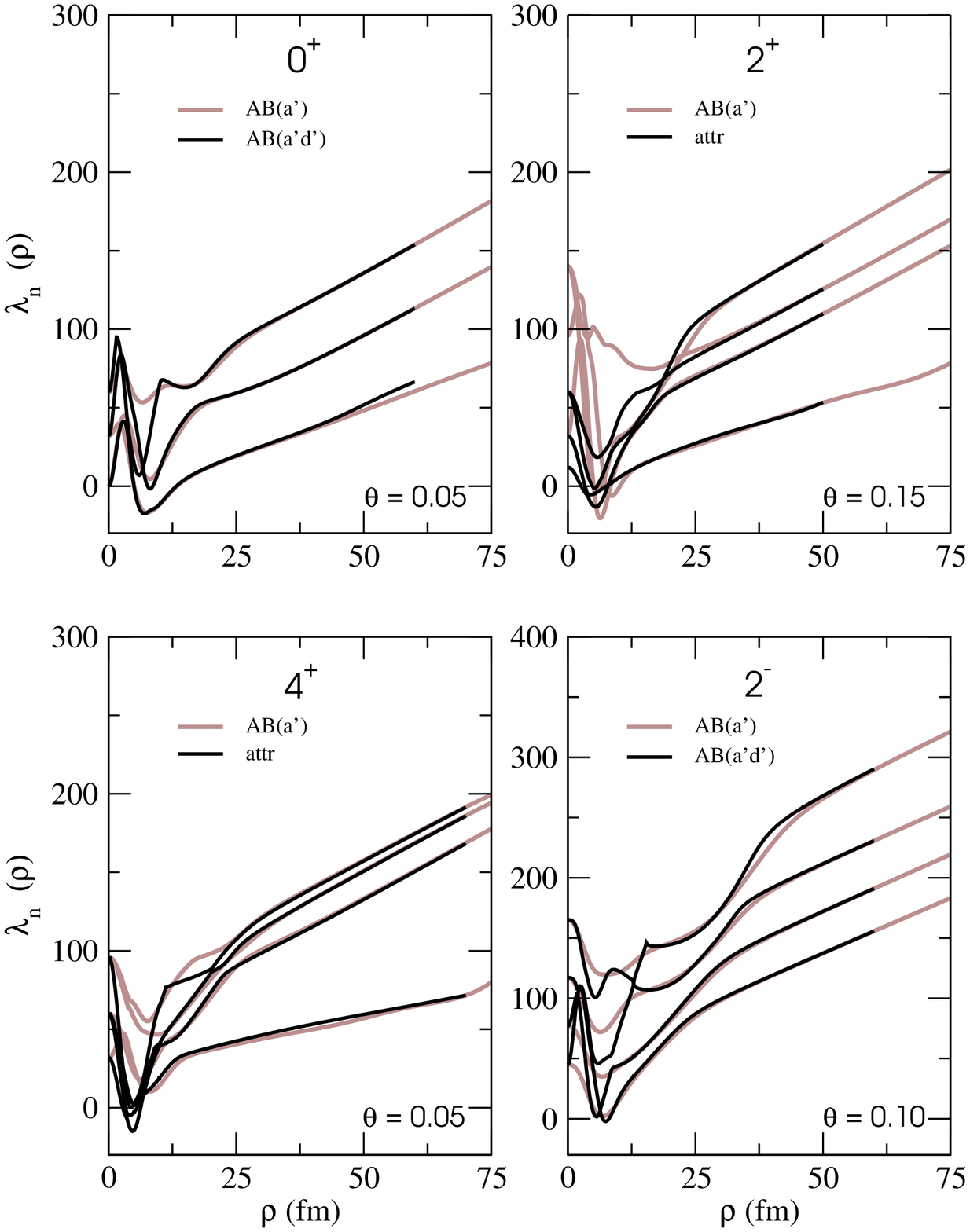,scale=0.75,angle=0}
\caption{The angular eigenvalues for some of the three-body resonances of 
$^{12}$C for different interactions. For the $0^{+}$ and $2^{-}$
three-body resonances we use ``AB(a')'' and ``AB(a'd')'', whereas we
use ``AB(a')'' and ``attr'' for the $2^{+}$ and $4^{+}$, see text for
specifications.  }
\label{eigen}
\end{figure*}

\subsection{The $3\alpha$-cluster model}

The model uncertainties are essentially related to the different
choices of interactions reproducing the low-energy $\alpha-\alpha$
scattering properties.  The Ali-Bodmer potentials with the correct
number of two-body bound states are constructed with repulsive cores
to simulate the effect of Pauli forbidden states. The radial
dependence then varies from being strongly repulsive to fairly
attractive implying rapidly varying adiabatic potentials with
(avoided) crossings. To test the sensitivity we constructed a purely
attractive potential ``attr'' with the two-body complex resonances
energies, $(E_R,\Gamma)= (100,0.01)$~keV, $(3.2,1.4)$~MeV, and
$(11.6,3.9)$~MeV in $\ell=0,2,4$ waves, respectively. These values are
rather close to the measured values.  We furthermore constructed the
potential ``AB(a'd')'' by combining the two-body $s$ and $d$-waves
from another Ali-Bodmer potential while substituting the $g$-wave by
the above attractive potential.  This potential has $\ell=0,2,4$
$^{8}$Be-resonances at $(E_R,\Gamma)= (90,0.006)$~keV,
$(3.0,1.4)$~MeV, and $(11.6,3.9)$ MeV, respectively.

The angular eigenvalues for different potentials are compared in
fig. \ref{eigen} for some of the resonances. The striking feature is
that the large-distance behavior of the spectra is the same.  First
the eigenvalues increase linearly with hyperradius simply because the
leading order is the Coulomb potential multiplied by $\rho^2$. Second
the coinciding spectra at large distances reflect that the resonance
positions are crucial. These in turn essentially only depend on the
two-body scattering lengths.  Both spectra and resonance energies are
to leading order then roughly model independent as determined by the
low-energy two-body scattering properties.

\begin{figure*}[h]
\epsfig{file=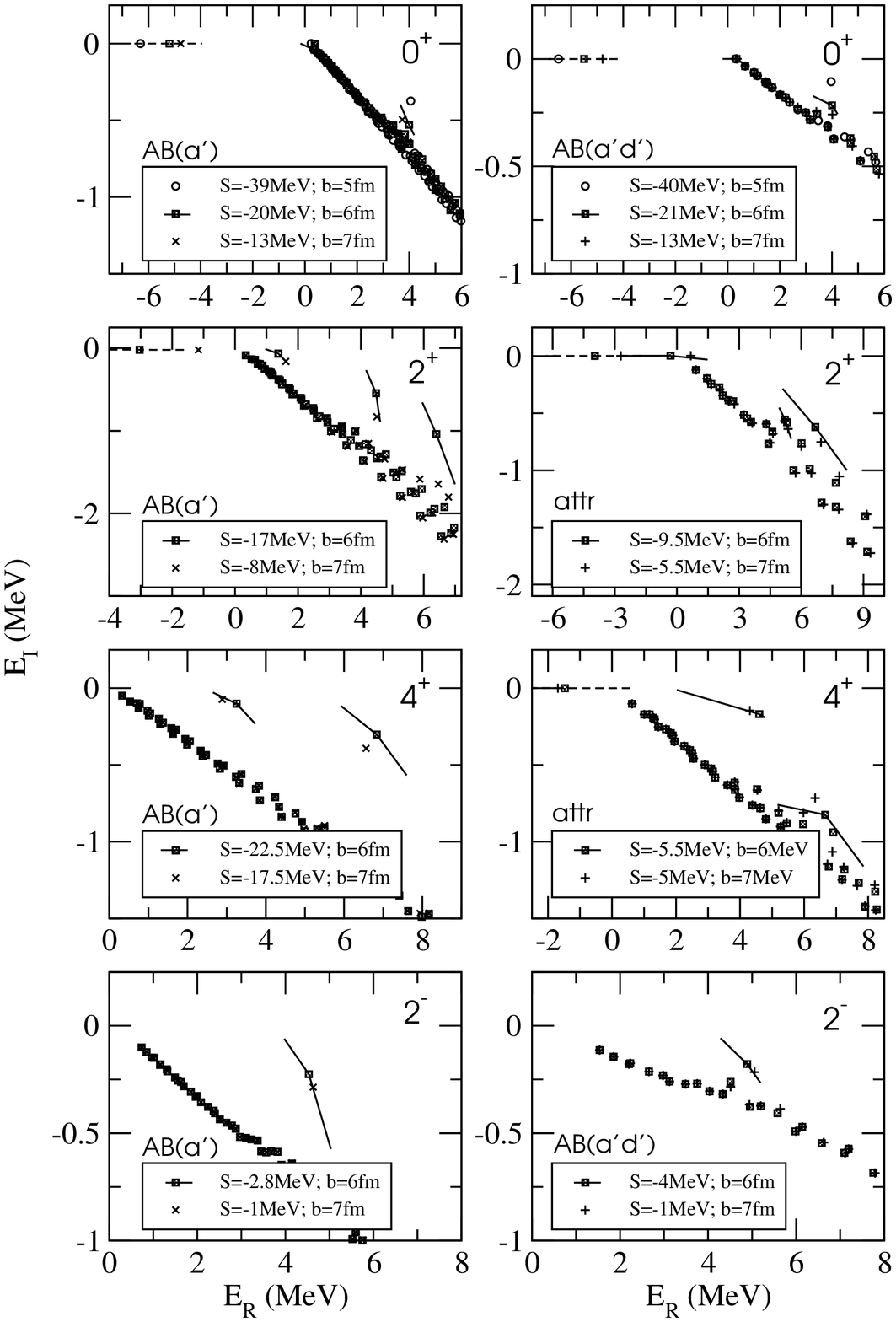,scale=0.75,angle=0}
\caption{The real and imaginary parts of the resonance energies for the 
four cases shown in fig. \ref{eigen} obtained after rotation by the
angle $\theta$ for different two-body interactions.  The left column
is for the two-body interactions ``AB(a')'' specified in
fig. \ref{potentials}.  The right column uses ``AB(a'd')'' for $0^+$
and $2^-$.  The three-body interactions specified by strength and
range are given in the figures.  The lines are the resonance positions
followed for $b=6$~fm when $S$ is varied $5$~MeV up or down from the
central value.  }
\label{spectr1}
\end{figure*}

The complex resonance energies are compared in fig. \ref{spectr1} for
different interactions.  First for the $0^+$-resonances we modestly
only change (from AB(a') to AB(a'd')) the $s$-waves a little while
both $d$ and $g$-potentials are quite different. The result is that
the energies essentially are unchanged for these potentials.  The
recently found experimental result has an unusually large width of
about 3.4~MeV \cite{dig05,dig06}. This is somewhat larger but still
comparable with the widths found from inelastic $\alpha-\alpha$
scattering experiments
\cite{joh03,ito04}. We then investigate the dependence on the
three-body interaction.  We follow the resonance as function of the
three-body strength for fixed values of the range. As the energy
increases also the width increases almost along the continuum
background spectrum.  If we instead keep the energy of the lowest
$0^+$-resonance by correlated changes of three-body strength and
range, the second $0^+$-resonance moves relatively little in position
while the width increases strongly. If both the measured position and
the unusually large width should be reproduced the range has to be
very large and the necessary rotation angle would not be numerically
accessible.

For the $2^-$-case we compare results from the same two interactions
where we find one resonance at roughly the same position and width in
both case.  This is more striking now where the dominating components
have $\ell_x = 2$ in contrast to $0^+$ where the similar $\ell_x =
0$-components dominate.  The variation with three-body strength is
rather similar.  The resonance position can easily be adjusted with
reasonable parameters in a fairly large interval around 5~MeV. Thus
even quite different potentials with the same two-body resonances give
roughly the same three-body structure.

For the $2^+$-resonances we compare in fig. \ref{spectr1} the results
for AB(a') with the results from, attr, a very different potential
with only attraction in all partial waves.  First we note that the
resonance positions vary differently with the three-body interaction
when we use the initial Ali-Bodmer potential.  The lowest only moves a
little and essentially parallel to the background continuum. The
second resonance has a strong tendency to move into the continuum
reflecting the quickly increasing width as a consequence of a
decreasing barrier. The third resonance moves faster but in the region
of the measured energy and width.

Second it is remarkable that the very different purely attractive
potential also leads to one bound state and three resonances in the
same energy region.  Again the lowest resonance is very close to the
$3\alpha$-threshold, and even below for a fairly weak attractive
three-body potential.  In this case it would seem natural to use a
repulsive three-body interaction but then the bound state is too
easily pushed up into the continuum. The level spacing is too small
as for most other cluster models.  The variation with three-body
potential is similar to the behavior seen for the other potential.
Now the second and third resonance positions are somewhat closer. From
these energy variations it is then again natural to associate the
highest-lying to the experimentally found resonance \cite{dig05,dig06},
and fix its position by adjusting the three-body strength for
different values of the range.  The width of this resonance and both
positions and widths of the other two resonances are then determined.

For the $4^+$-resonances and the initial Ali-Bodmer potential we
notice the same behavior as for the $2^+$-resonances, i.e. the lowest
tends to increase the width quickly and disappear into the background
whereas the highest moves faster with the three-body strength in the
region of the measured energy and width. This is again due to the
structure of the potential.  The resonance energy in the minimum is
pushed up towards the top of the barrier where the width increases
dramatically.  The highest $4^+$-resonance can then, in contrast to
the lowest, rather easily be moved to the experimental position by
adjusting the three-body strength.

For the purely attractive potential we have to choose a repulsive
three-body interaction if no three-body bound states should
appear. With the same range of $b=6$~fm the dominating potential has a
high and thick barrier leading to a very small width.  Instead of an
arbitrary adjustment of the range for this particular case we use the
same ranges and a similar attractive three-body potential. The $4^+$
bound state in fig.~\ref{spectr1} should then be considered spurious
and discarded.  Then again two resonances appear with energies below
8~MeV and widths varying from almost zero to almost 2~MeV.  This is
within a factor of two from the results of the initial Ali-Bodmer
potential.

\begin{figure*}[h]
\epsfig{file=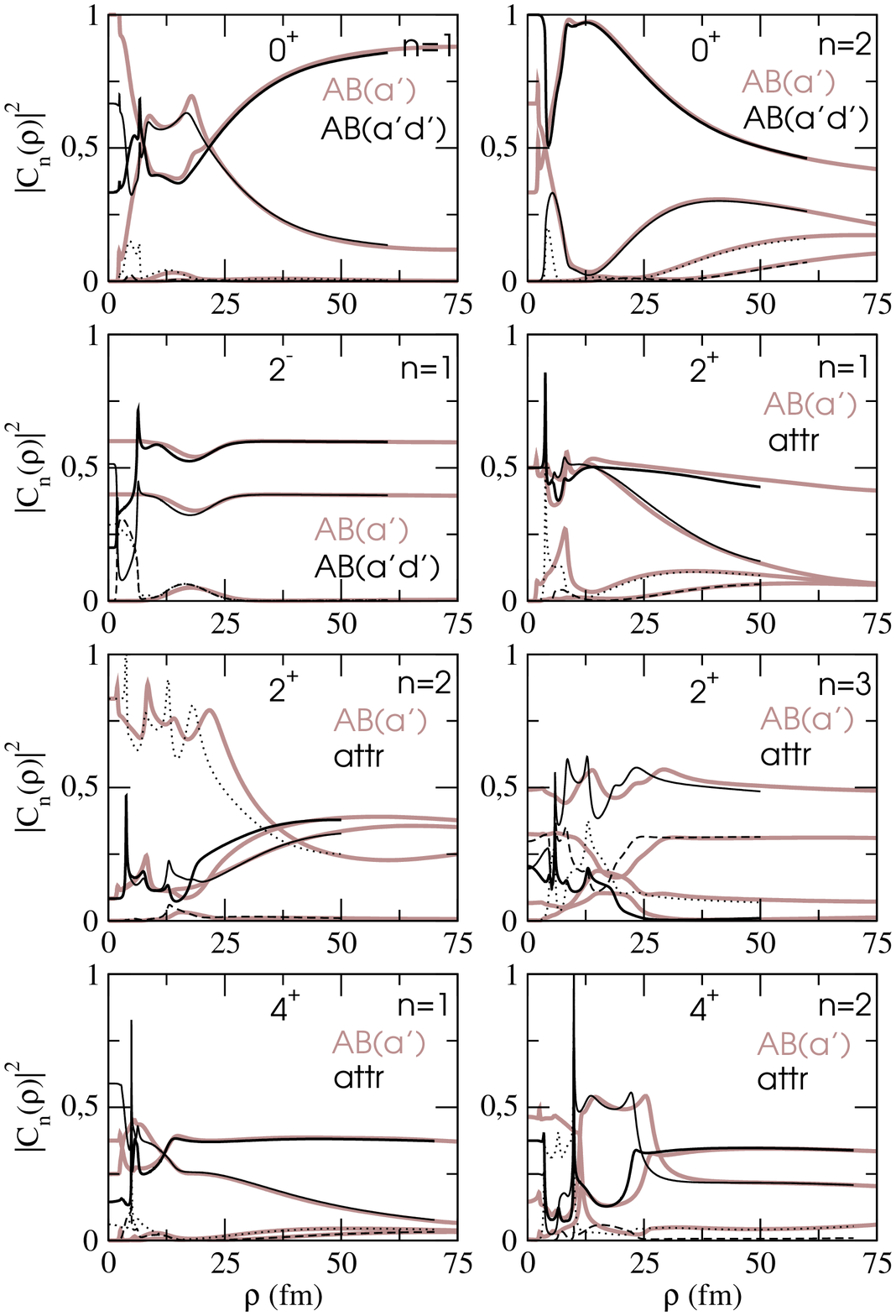,scale=0.75,angle=0}
\caption{The partial wave decomposition of the same $^{12}$C  
resonances for the two-body interactions given in the figures. }
\label{contra1}
\end{figure*}

The structures of the resonances are reflected in their partial wave
decompositions.  The large-distance properties of the eigenvalues are
almost identical for interactions with the same two-body resonances,
and consequently the partial wave components also coincide for large
hyperradii.  However, it is also obvious that when the short-distance
properties are qualitatively different then the adiabatic potentials
must reflect those differences. 

We compare the quantitative differences in fig. \ref{contra1} for some
of the adiabatic potentials.  For $0^+$-states the $s$ and $d$-waves
dominate and the change into an attractive $g$-wave has only a small
effect except at very small hyperradii.  For the $2^+$-states we
compare results where all partial wave two-body interactions are
different.  Nevertheless, the partial wave decompositions for the
three lowest adiabatic potentials are remarkably unaffected.  The same
behavior of invariant large-distance properties and relatively little
variation at small distance is found for the $4^+$ and $2^-$-states.

The other resonances are all easily moved to the desired position by
the three-body interaction, and the widths are strongly correlated
with the measured values, as in table \ref{tab2}. The structures of
these resonances are also very insensitive to the specifics of the
interactions.  The $6^+$-resonance seems consistently to fall at an
energy of about 6~MeV which is in the upper end of the considered
energy interval.  This estimate has about the same uncertainty as the
other natural parity even angular momenta states.  They are all built
on the same low angular momentum partial waves, see table \ref{tab1}.
The additional uncertainty is due to the less known two-body
interactions of higher angular momentum.  The $3^+$-resonance is much
more uncertain since the necessary three-body potential could vary
from almost vanishing to values similar to that of $1^+$.

We can now summarize how resonance energies and structures are
influenced by two-body interactions reproducing the low-energy
scattering properties.  First the number of resonances remain
unchanged within approximately the same region of complex energies.
Their energies may vary by modest amounts. One exception is the
$4^+$-states with the purely attractive potential where the bound
state has to be discarded as Pauli forbidden.  The structures of the
adiabatic potentials are rather insensitive to the specific
interaction except at small hyperradii corresponding to distances
where the three $\alpha$-particles overlap substantially.  The
relative contributions of the adiabatic potentials may vary at small
distances where the contributions to the resonances are rather small.

In particular, the insensitive properties are that (i) the bound and
the lowest $0^+$-states are related to the same adiabatic potential
differing by one radial oscillation, (ii) the bound and the second
$2^+$-states also are related to the same adiabatic potential
differing by one radial oscillation, (iii) the second $0^+$-resonance
and the first (bound) $2^+$ state have no radial nodes and both are
related to adiabatic potentials different from those associated to the
lowest resonances, (iv) the third $2^+$-state has no radial nodes and
it is related to a third adiabatic potential, (v) all states only
appearing once are related to the lowest adiabatic potential without
radial nodes.

\subsection{Other models}

All angular momenta and parities are also considered in some previous
model computations. The basic problem in comparison is that the
interactions often are different in the many investigations.  However,
the parameters are always chosen to reproduce some measured
quantities. The shortcoming of such an approach is in general that
misinterpreted data inadvertently can be used both to fit the
interactions and afterwords to compare to model results.  To
illustrate the variation of the model predictions we discuss some
results from selected representative methods which all to some degree
include the underlying microscopic nuclear structure.

The formulation from antisymmetrized molecular dynamics \cite{kan98}
seems first to aim at reproducing the bound state properties. The
low-energy excited spectrum then differs from ours only by one less
$2^+$-resonance.  The level spacing is a little larger, and the $1^+$
and the last $4^+$-states are found close to 17.5~MeV.  We find all
our states below 16~MeV and the spacing between the two bound states
too small.  Our philosophy is that the ground state has contributions
beyond the $3\alpha$-cluster structure and we should not necessarily
reproduce its energy.  In the newest preprint \cite{kan98} the
spectrum has changed reflecting the use of different interactions.
Now the number of resonances with given angular momentum and parity
below 16~MeV are the same as in the present model.

The method of fermionic molecular dynamics is less developed in
practical applications \cite{nef04}.  They find one less $2^+$, $1^+$
and $4^-$-resonance than us below 15~MeV.

An older publication using the generator coordinate method found many
resonances below 15~MeV \cite{ueg77,ueg79}.  They found one less $2^+$
than us, and otherwise the same number of states in this spectrum.
The generator coordinate method has recently been used extensively to
investigate the structure of light nuclei \cite{des02}.  The focus is
here on the bound states and the lowest resonances.  The two bound
states are too close-lying, and the number of resonances are too few
both compared to experiments and to our results. Below 15~MeV they
find only one $0^+$, no $1^+$, one $2^+$, and no $2^-$.  Furthermore,
their sequence has very little resemblance with the measured spectrum.

The algebraic cluster model is based on group theory \cite{bij99}.
Two more states have appeared in the last publication close to an
excitation energy of 15~MeV.  Below 15~MeV they find the same spectrum
as us except for one less $4^+$-state. The overall spacing is
therefore almost the same.  The spectrum is computed from a model with
3 $\alpha$-particles in an equilateral triangle where
rotation-vibration interactions, Coriolis forces and
vibration-vibration interactions are needed.

\subsection{Experiments}

In confronting the results in table \ref{tab2} with experimental input
the predicted low-lying 2$^+$-states and the lowest 4$^+$-state stand
out as problematic. However, experimentally there is not presently a
consensus on the position of 2$^+$-resonances in $^{12}$C; as
previously mentioned different probes find states at different
energies. To some degree this might simply reflect selection rules or
structural effects in different experiments, but when the same probe
sees different states in different experiments (as for
$^{12}$C($\alpha$,$\alpha$)$^{12}$C~\cite{joh03,ito04}), there is
clearly a problem. It seems safe to infer from the data that a broad
2$^+$-state exists in the region 14-15 MeV as e.g. seen in
$\beta$-decay \cite{dig05,dig06} and in some scattering
experiments\cite{ajz90}, but whether more broad states exist near
10~MeV overlapping with the known broad 0$^+$-state, as suggested by
\cite{ito04}, is unclear. If so they must be feebly fed in the
$\beta$-decays as suggested in \cite{kan98}. The large width of the
second $0^+$-resonance is qualitatively in agreement with the results
of about $2.7$~MeV obtained from the inelastic scattering experiments
\cite{joh03,ito04}.  The existence of narrow states below 10~MeV can 
be ruled out experimentally. The existence of a broad 4$^+$-state
below the known one at 14.1~MeV cannot be ruled out experimentally;
for comparison the broad 0$^+$-state centered at 10~MeV is not seen in
a number of transfer reactions due to its large width $-$ it is most
clearly seen in $\beta$-decay experiments where a 4$^+$-state cannot
be favorably populated.

\section{Summary and conclusions}

We employ the established method of hyperspherical adiabatic expansion
in combination with complex scaling to compute the energies and widths
of $^{12}$C-resonances below 15.96~MeV.  We use the $3\alpha$-cluster
model with well known angular momentum dependent two-body interactions
adjusted to reproduce the low-energy $\alpha-\alpha$ scattering
properties. Obviously then only three-body properties can be computed,
but within this overall constraint, the model can be applicable even
when many-body properties are important.  The partial waves of each
contributing Faddeev component are expanded on a large complete basis
of Jacobi polynomials. Only very few adiabatic potentials are needed
for convergence for each angular momentum and parity $J^{\pi}$.

The energies are determined by the behavior of the potentials at small
distances. At intermediate distances the potential has a barrier that
determines the resonance width.  As mentioned, the two-body potential
is chosen to reproduce the low-energy two-body scattering properties,
and a three-body short-range interaction is also added. This accounts
for the intrinsic nucleonic degrees of freedom by producing a correct
boundary condition at small distances matching the $3\alpha$-structure
at large distance.  However, it does not describe the short-distance
structure properly and hence electromagnetic transitions can not be
expected reliably estimated in this model.

In our three-body cluster model we add three-body potentials to place
the resonances at the desired positions.  These potentials are chosen
independently for each $J^{\pi}$ to depend only on the hyperradius.
We use one Gaussian with a range $b=6$~fm corresponding to the
distance when three $\alpha$-particles touch each other in an
equilateral triangle. The strength is used to place one of the
resonances at the desired energy.  This invention opens the cluster
model for applications to many-body resonances similar to the simple
$\alpha$-emission model.

We are not concerned with the bound states where especially the ground
state can deviate more from a $3\alpha$ cluster state than most of the
excited states and here in particular from the first $0^+$ resonance.
We focus instead on reproducing the correct energy and investigate the
systematics of the strongly correlated corresponding width. Designing
a three-body potential to fit simultaneously more of the states would
not provide any more insight.  The resonance structures are
essentially independent of these three-body potentials.

The model leads to a number of low-lying resonances for each
$J^{\pi}$, i.e. two $0^{+}$, three $2^{+}$, two $4^{+}$, and one of
each of $1^{\pm}$, $2^{-}$, $3^{\pm}$, $4^{-}$, and $6^{+}$.
Dependence on the two and three-body interactions are investigated.
The conclusions are that the highest of the $2^{+}$ and $4^{+}$
resonances most naturally can be placed at the measured values whereas
the lowest of these resonances acquire a very large width when they
are pushed towards higher energies. In fact the confining barriers of
the adiabatic potentials vanish before the energy is high enough.

The structures of the individual resonances are almost totally
independent of the three-body potential or equivalently independent of
the resonance positions.  On the other hand, the energies and widths
are strongly correlated and rather insensitive to the structure. With
the AB(a') Ali-Bodmer potential and the measured energies reproduced
with a three-body potential of range for $b=6$~fm, the widths are all
systematically about a factor of two larger than measured indicating a
corresponding spectroscopic factor of about 0.5.  Two exceptions
appear, i.e. the second $0^{+}$ resonance with a computed width about
3 times smaller than a recent measurement, and the two $1^{+}$
resonances with $3-5$ orders of magnitude larger computed widths and
the corresponding very small spectroscopic factors.

One of each of the negative parity states is found suggesting that the
tentative assignment of one of the $2^-$-resonances should be changed
to $4^-$.  The large computed width of the second $0^{+}$-resonance
appears in the same energy region as several other resonances.
The two lowest $2^{+}$ and the lowest $4^{+}$ resonances are
presumably impossible to reconcile with the experimental information.
The energies of both the lowest $2^{+}$ and $4^{+}$-resonances should
presumably be higher.  This can be achieved by using a three-body
interaction depending on more than the hyperradius. However, in the
present work the focus is on the structure and the relation between
energies and widths.

The simple three-body cluster model is, perhaps surprisingly, doing as
well as more microscopic models. The difference is really that the
phenomenology is inserted on the nucleonic level for microscopic
models.  These microscopic models are more likely to have problems
with spatially extended systems, and elaborate calculations of
observables. Our cluster model exploits the experimental information
by inserting the phenomenology on the $\alpha$ cluster level and a few
of the $^{12}$C properties are used as well.  This model is still
technically difficult but closer to the observables since we adjust on
the three-body level without structure changes.

The discrepancies between experimental and theoretical results can
logically be related to either experiments or theory or to a
combination of both.  Cleaning up on the experimental side to achieve
consistency between different experiments is necessary.  This is
especially in connection with the possibly overlapping resonances
around 4~MeV.  The present model does not treat the intrinsic degrees of
freedom.  They are accounted for on the two-body level by
phenomenologically adjusted interactions, and on the three-body level
by the hyperradial and $J^{\pi}$ dependent potential. The most likely
suspect is the simplicity of the three-body interaction which is
supposed to mock up the important effects of the nucleonic degrees of
freedom.  The Pauli exclusion, when three $\alpha$-particles are
close, is perhaps not well described.  The one-Gaussian structure may
be too simple, and the potential should perhaps depend on angular
momentum quantum numbers or on individual adiabatic potentials.

In conclusion, we find in the present model a comparable number of
resonances as in more microscopic calculations but the detailed energy
spectra vary substantially between the different models.  We find more
low-lying resonances than measured.  The partial wave structures of
our resonances are very robust independent of their energies. We have
established resonance structures and suggested the most likely
energies within the model.

The real test is then to compare experiment and theory for observables
with strong structure dependence.  This would demonstrate which of the
computed resonances are really seen in experiments. The obvious, and
perhaps the only, observable for this is the momentum distributions of
the $\alpha$-particles after decay of the resonances.  Experimentally
this has to be both accurate and kinematically complete. Theoretically
large-distance continuum properties must be accurately computed and
the present technique is at the moment the closest to provide answers.
Both the necessary experiments and calculations are challenging but
possible and both within reach.

\begin{center}
{\Large \bf Acknowledgments} 
\end{center}

One of us ( R.A.R.) acknowledges support by a post-doctoral fellowship from
Ministerio de Educaci\'on y Ciencia (Spain).

\end{document}